\documentclass[aps,prd,onecolumn,groupedaddress,nofootinbib,showpacs,showkeys,amssymb]{revtex4}
\usepackage[colorlinks,citecolor=blue,urlcolor=blue,linkcolor=blue]{hyperref}
\newlength{\extraspace}
\newlength{\extraspaces}
\begingroup
\endgroup

\newcommand{\be}{\begin{equation}
\addtolength{\abovedisplayskip}{\extraspaces}
\addtolength{\belowdisplayskip}{\extraspaces}
\addtolength{\abovedisplayshortskip}{\extraspace}
\addtolength{\belowdisplayshortskip}{\extraspace}}
\newcommand{\ee}{\end{equation}}

\newcommand{\ba}{\begin{eqnarray}
\addtolength{\abovedisplayskip}{\extraspaces}
\addtolength{\belowdisplayskip}{\extraspaces}
\addtolength{\abovedisplayshortskip}{\extraspace}
\addtolength{\belowdisplayshortskip}{\extraspace}}
\newcommand{\ea}{\end{eqnarray}}

\begin{document}
\title{Spherically symmetric  charged black hole in  conformal teleparallel equivalent of general relativity}
\author{G.G.L. Nashed$^{1,2,3}$}
\email[Electronic address: ]{nashed@bue.edu.eg}
\author{Kazuharu Bamba$^{4}$}%
\email[Electronic address: ]{bamba@sss.fukushima-u.ac.jp}
\affiliation{$^{1}$Centre for theoretical physics, the British University in Egypt, 11837 - P.O. Box 43, Egypt}
\affiliation{$^{2}$Mathematics Department, Faculty of Science, Ain Shams University, Cairo, Egypt}
\affiliation{$^{3}$Egyptian Relativity Group (ERG)}
\affiliation{$^{4}$Division of Human Support System, Faculty of Symbiotic Systems Science, Fukushima University, Fukushima 960-1296, Japan}

\begin{abstract}
 We studied 4-dimensional non-charged and charged  spherically symmetric spacetimes in conformal teleparallel equivalent of general relativity.  For this aim, we apply the field equations of non-charged and charged to diagonal and non-diagonal vierbeins and derive their sets of non-linear differential equations. We investigate in details that the Schwarzschild, for the non-charged case, and  the Reissner-Nordstr\"om, for the charged case,  are the only black hole solutions for the spherically symmetric case in the frame of conformal teleparallel equivalent of general relativity theory. Our conclusion indicates that the scalar field in the conformal teleparallel equivalent of general relativity  theory has no effect for the spherically symmetric manifold. \vspace{0.2cm}\\
\end{abstract}

\maketitle

\section{Introduction}\label{S1}
The conformal symmetry was introduced by Deser,  Dirac and Utyiama \cite{Ds,Dp,Ur} as a basic   construction of the Einstein  General Relativity (GR) theory.  Dirac constructed the conformal invariant procedure to the Einstein  theory  \cite{Dp} as a novel variational principle for the action of GR introducing a scalar  field, beside the metric components  $g_{\mu \nu}$ \cite{PA}.  The dealing of the conformal gravity was supported by the theory of Ogievetsky \cite{Ov} which states that  {\it  ``general relativity-diffeomorphism group can be obtained as the closure of two finite-dimensional groups"}.

Conformal symmetry is an important topic  to quantize
gravity \cite{Hg}. This symmetry is recognized in modified  constructions
of GR, which amendment  the small and large scale descriptions
of the spacetime. In the
small distances one has to  modify the  behavior of the gravitational field  to formulate a renormalizable and
a unitary  quantum gravity theory. At the large scale domains, the modified theories have
to give a consistent solution to the dark energy and dark matter problems. It is shown that  a conformal transformation can be carried
 out in the non-conformally invariant of the Einstein- Hilbert action so that the conformal factor $\Omega(x)$ can be pull out
from the metric tensor $g_{\mu \nu}$ according to the formula $g_{\mu \nu} =\Omega(x) \hat{g}_{\mu \nu}$, which can be  thought
as extra degree of freedom that we can deal with it as an independent variable \cite{Hg}. The quantity $\hat{g}_{\mu \nu}$ is not a tensor due to the fact
that $det \hat{g}=-1$  however, one can show that the effective theory constructed out of $\hat{g}_{\mu \nu}$ is a scale
invariant \cite{Hg10}.  The conformally invariant Weyl theory, that is quadratic
in curvature tensor, have been extensively studied in the literature \cite{Mp,MS}.

Einstein's GR is not the only gravitational theory, there is also another gravitational theory, ``Teleparallel Equivalent of General Relativity (TEGR)" that is introduced in the literature by Einstein \cite{Ea28} (for recent reviews on modified gravity theories to explain the so-called dark energy problem, see, e.g.,~\cite{Nojiri:2010wj, Capozziello:2011et, Nojiri:2017ncd, Capozziello:2010zz, Bamba:2015uma, Bamba:2012cp}). In this theory, the gravitational field is described by the torsion tensor unlike GR whose gravitational field is described by curvature \cite{PP,HS,Mj1}. The main motivation of TEGR is the fact that within its frame one can define a consistent expression for energy, momentum and angular-momentum for the gravitational system \cite{M1,MUFN,Mj}.

In the framework of conformal transformation there are two methods  to write the gravitational Lagrangian: The first method  is the one in which we add a scalar field to the Einstein-Hilbert action; while in the second method one has to choose the Lagrangian which is a quadratic term of Weyl tensor. Maluf et al. \cite{MF} and Silva et al. \cite{SSU} have used the first method to construct a conformal teleparallel  theory. Silva et al. \cite{SSU}, in the framework of conformal TEGR, have shown that for a flat  homogenous isotropic spacetime there is an explanation for accelerated and expanded universe. It is the purpose of the present study to apply the field equation of conformal TEGR theory to a 4-dimension spherically symmetric spacetime and see what is the effect of the scalar field. 
We note that the proof of the Birkhoff-like theorem has been proposed for spherical, vacuum, asymptotically flat Brans-Dicke/scalar-tensor gravity~\cite{Hawking:1972qk, Mayo:1996mv, Bekenstein:1996pn, Sotiriou:2011dz, Bhattacharya:2015iha}.

The structure of this study is as follows. In Section \ref{S1.1}, we briefly review the conformal teleparallel equivalent of general relativity (CTEGR)  formalism using the  tensors definitions and the field equations. In  Section \ref{S2}, a diagonal $4$-dimension  vierbein field having spherical symmetry  is applied to the  field equations of CTEGR gravity to obtain a general neutral black hole solution in 4-dimensions, that is asymptotically flat.  In Section \ref{S3}, a non-diagonal spherically symmetric vierbein  is applied to the CTEGR gravity and a solution similar to the diagonal case is derived.  In Section \ref{S4}, we derived the charged field equations of CTEGR. Also, in Section \ref{S4}, we apply the charged field equations to the diagonal vierbein field that used in the static case. A general charged solution which is the only physical solution is derived  in Section \ref{S4}.  We repeat our terminology to the non-diagonal spherically symmetric spacetime and solution similar to the diagonal case   is derived in section \ref{S5}. A conclusion of our results is  reported in Section \ref{S6}.

\section{The conformal of teleparallel equivalent of general relativity}\label{S1.1}
The teleparallel equivalent of general relativity (TEGR)  is prescribed by\footnote{In this study we will use the Latin indices ${\it i, j, \cdots }$ for the tangent spacetime coordinates and the Greek ones $\alpha$, $\beta$,
$\cdots$  to label the (co)-frame components.}   $\{{\it M},~L_{i}\}$, where $\it M$  is the $4$-dimensional spacetime and $L_{i}$ ($i=1,2,3,4$) are the  vectors that defined globally on the spacetime $\it M$. The vector fields  $L_{i}$ are considered as the parallel  vectors. In the $4$-dimension, the parallel  vectors  are coined as  the {\it vierbein or tetrad} fields and their   contravariant   derivative are vanishing i.e.,
\begin{equation}\label{q1}
  D_{\mu} {L_i}^\nu:=\partial_{\mu}
{L_i}^\nu+{\Gamma^\nu}_{\lambda \mu} {L_i}^\lambda= 0,
\end{equation}
where   differentiation is carried out with respect to the non-symmetric affine connection which is called the  Weitzenb\"{o}ck connection,    ${\Gamma^\nu}_{\lambda \mu}$, which is defined as   \cite{Wr}
\begin{equation}\label{q2}
{\it {\Gamma^\lambda}_{\mu \nu} := {L_i}^\lambda~ \partial_\nu L^{i}{_{\mu}}},
\end{equation}
and  \[\partial_{\nu}:=\frac{\partial}{\partial x^{\nu}}.\]
The metric tensor in this geometry is defined as
\begin{equation}\label{q3}
 {\rm g_{\mu \nu} :=  \eta_{i j} {L^i}_\mu {L^j}_\nu},
\end{equation}
with $\eta_{i j}=(+,-,-,- )$ being  the {\cal 4}-dimensional Minkowski metric. The metricity condition is satisfied  as a result of Eq. (\ref{q1}). The torsion, ${\mathrm{T}^\alpha}_{\mu \nu}$  and the contortion, $\mathrm{K}^{\mu \nu}{}_\alpha$ tensors are defined as
\begin{eqnarray} \label{q33}
\nonumber {T^\alpha}_{\mu \nu}  & := &
{\Gamma^\alpha}_{\nu \mu}-{\Gamma^\alpha}_{\mu \nu} ={L_i}^\alpha
\left(\partial_\mu{L^i}_\nu-\partial_\nu{L^i}_\mu\right),\\
{K^{\mu \nu}}_\alpha  & := &
-\frac{1}{2}\left({T^{\mu \nu}}_\alpha-{T^{\nu
\mu}}_\alpha-{T_\alpha}^{\mu \nu}\right). \label{q4}
\end{eqnarray}
The vector of the torsion (contraction of the torsion) is defined as
\begin{equation}\label{Tv}
{\rm T_\nu := {T^\mu}_{\mu \nu}}.
\end{equation}
The scalar of torsion of the TEGR theory  is defined as
\begin{equation}\label{Tor_sc}
{\rm T := {T^\alpha}_{\mu \nu} {S_\alpha}^{\mu \nu}},
\end{equation}
with ${S_\alpha}^{\mu \nu}$ being the superpotenial tensor which is  skew symmetric tensor   in its first pair and is defined as
\begin{equation}\label{q5}
{\rm {S_\alpha}^{\mu \nu} := \frac{1}{2}\left({K^{\mu\nu}}_\alpha+\delta^\mu_\alpha{T^{\beta
\nu}}_\beta-\delta^\nu_\alpha{T^{\beta \mu}}_\beta\right)}.\end{equation}
The Lagrangian of the TEGR theory is given by
\begin{equation}\label{q6}
\mathcal{L}({L^i}_\mu)_{g}=\frac{|L|T}{2\kappa}, \qquad \qquad \textrm {where} \qquad \qquad L=\sqrt{-g},
\end{equation}
with $\kappa=8\pi$ being the coupling gravitational constant.

The Lagrangian (\ref{q6}) is not invariant under the conformal transformations
since \begin{equation}\label{q77} \bar{L}_{a \mu}=e^{\omega(x)} L_{a \mu},\end{equation} where $\omega(x)$ is an arbitrary. Maluf et al.  \cite{MF} have investigated that Lagrangian (\ref{q6}) in not invariant under the  transformation (\ref{q7}). Therefore, we should modify Eq. (\ref{q6}) so that it becomes invariant under the transformation (\ref{q77}). It was shown that the Lagrangian which is invariant under conformal transformation is given by \cite{MF}
\begin{equation}\label{q7}
\mathcal{L}({L^i}_\mu, \phi)_{g}=2\kappa|L|\left[-\phi^2T+6g^{\mu \nu}\partial_\mu \phi\partial_\nu \phi-4g^{\mu \nu}\phi(\partial_\nu \phi) T_\mu\right]+\mathcal{L}_m,
\end{equation}
where $\phi$ is a scalar field and $\mathcal{L}_m$ is the Lagrangian of the matter fields. The transformation of the scalar field
\[ \phi \rightarrow {\bar \phi}=e^{-\omega(x)} \phi,\] which makes the Lagrangian (\ref{q7}) invariant.
Making the variational principle to the Lagrangian (\ref{q7}) with respect to the scalar field $\phi$ one can get \cite{MF}
\begin{equation}\label{q8}
I\equiv \partial_\mu(Lg^{\mu \nu} \partial_\nu \phi)-\frac{L}{6}\phi R-\frac{\kappa}{6}\frac{\delta \mathcal{L}_m}{\delta \phi}=0,\end{equation}  where the Ricci scalar $R$ can be rewritten as $$LR=2\partial_\nu(LT^\nu)-LT.$$

By the same method if one carry out the variational principle of the Lagrangian density (\ref{q7}) with respect to the vierbein $L_{a \mu}$ one can get \cite{MF}
\begin{eqnarray}\label{q9}
&&Q^{a \nu}\equiv \partial_\alpha(L\phi^2 S^{a \nu \alpha})-L\phi^2(S^{b  \alpha \nu} T_{b \alpha}{}^a-\frac{1}{4}L^{a \nu} T)
 -\frac{3}{2}LL^{a \nu}g^{\beta \mu}\partial_\beta \phi \partial_\mu \phi+3LL^{a \mu}g^{\beta \nu}\partial_\beta \phi \partial_\mu \phi
 +LL^{a \nu}g^{\beta \mu}  T_\mu \phi \partial_\beta \phi\nonumber\\
& & -L\phi L^{a \beta}g^{\nu \mu} ( T_\mu  \partial_\beta \phi+T_\beta  \partial_\mu \phi)
 -Lg^{\beta \mu}  \phi T^{\nu a}{}_\mu \partial_\beta \phi-\partial_\mu[L g^{\beta \nu} L^{a \mu} \phi \partial_\beta \phi]+\partial_\rho[L g^{\beta \rho} L^{a \nu} \phi \partial_\beta \phi]-\frac{\kappa}{2}\frac{\delta \mathcal{L}_m}{\delta L{a \nu}}=0,\end{eqnarray}
where $$\frac{\delta \mathcal{L}_m}{\delta L_{a \nu}}=LL^a{}_\mu {\mathop{\mathfrak{T}}}^{\mu \nu}.$$

It is important to stress on the fact that the field equation (\ref{q9}) is reduce to the field equation of TEGR when the scalar field $\phi=1$ \cite{AP}--\cite{MDTC}. It is also important to note that the trace of Eq. (\ref{q8}) is given by
\[\phi \frac{\delta \mathcal{L}_m}{\delta \phi}=L\mathop{\mathfrak{T}},\]
with\begin{equation} \mathop{\mathfrak{T}}=g_{\nu \mu} {\mathop{\mathfrak{T}}}^{\nu \mu},\end{equation} being the trace of the energy-momentum tensor. Therefore, for the traceless energy momentum tensor the term $\frac{\delta \mathcal{L}_m}{\delta \phi}$ is vanishing identically. In the following sections we are going to study the effect of the scalar field $\phi$ on the spherically symmetric spacetime.

\section{Spherically symmetric black hole solution: Diagonal vierbein}\label{S2}
We apply the  field equations of CTEGR given by Eqs. (\ref{q8}) and  (\ref{q9}) to the spherically symmetric $4$-dimensional  spacetime, which directly gives rise to the diagonal vierbein  written in the polar  coordinate ($r$, $\theta$, $\phi$, $t$) as follows \cite{,N008,N010,Gw10}:
\begin{equation}\label{vierbein}
\hspace{-0.3cm}\begin{tabular}{l}
  $\left({L_{i}}^{\mu}\right)=\left(\frac{1}{\sqrt{N(r)}}, \; r, \; r\sin\theta\; , \sqrt{K(r)}\;, \right)$
\end{tabular}
\end{equation}
where $N(r)$ and $K(r)$ are two unknown functions of the radial coordinate $r$. The metric associated with vierbein (\ref{vierbein}) has the form
 \begin{equation}\label{met} ds^2=K(r) dt^2-\frac{dr^2}{N(r)}-d\Omega^2, \qquad \textit{where} \qquad d\Omega^2=r^2(d\theta^2+sin^2\theta d\phi^2).\end{equation}
 Using Eq. (\ref{vierbein}) into Eq. (\ref{q33}) we get the non-vanishing components of the torsion $T^{a b c}$,  the vector of the torsion and contorsion $K^{a b c}$ in the form\footnote{The components of the torsion  $T^{a b c}$ is defined as \[T^{a b c}=L^a{}_\mu L^b{}_\nu L^c{}_\alpha T^{\mu \nu \alpha}.\] Same definition can be applied for the tensors $K^{\mu \nu \alpha}$ and $S^{\mu \nu \alpha}$. It
is important to recall  that the torsion and contorsion components are skew-symmetric in the last two indices while the contorsion is skew symmetric in its first two ones }:
\begin{eqnarray}\label{TC1}
 && T^{(2) (2) (1)}=T^{(3) (3) (1)}=\frac{\sqrt{N}}{r}, \qquad  \qquad T^{(4) (1) (4)}=\frac{\sqrt{N}K'}{2K}, \qquad \qquad  T^{(3) (3) (2)}=\frac{\cot\theta}{r}, \nonumber\\
& & T_{(1)}=-\frac{4K+rK'}{2rK}, \qquad T_{2}=-\cot \theta,\nonumber\\
& & K^{(2) (1) (2)}=K^{(3) (1)  (3)}=\frac{\sqrt{N}}{r}, \qquad \qquad  K^{(1) (4) (4) }= \frac{\sqrt{N}K'}{2K},\qquad \qquad K^{(3) (2) (3) }= \frac{\cot\theta}{r}.
\end{eqnarray}
Using Eq. (\ref{TC1}) we get the non-vanishing components of the superpotential in the form:
\begin{eqnarray}\label{sup}
 && S^{ (1) (2)  (1)}=S^{ (4) (4) (2)}=\frac{\cot\theta}{2r}, \qquad  \qquad S^{(4) (4) (1)}= \frac{\sqrt{N}}{r},\qquad \qquad  S^{(2) (1) (2)  }=S^{(3) (1) (3)  }= \frac{\sqrt{N}(2K+rK')}{4rK}.
\end{eqnarray}
Substituting from Eqs. (\ref{TC1}) and (\ref{sup}) into Eq. (\ref{Tor_sc}), we evaluate the torsion and the Ricci scalars as
\begin{equation}\label{df1}
T=\frac{2N(K+rK')}{r^2K},\qquad \qquad R=\frac{4rK^2N'-r^2NK'^2+2r^2NKK''+r^2KN'K'+4NK^2-4K^2+4rKNK'}{r^2K^2}.
\end{equation}
Using Eqs. (\ref{TC1}), (\ref{sup}) and (\ref{df1}) in  the field equations (\ref{q7}) and (\ref{q8}) when ${\mathop{\mathfrak{T}}}^{\nu \mu}=0$ we get the following non-vanishing components:
\begin{eqnarray}\label{df11}
& & Q^{(1)}{}_r\equiv 3r^2NK\Phi'^2+4rNK\Phi \Phi'+r^2N\Phi K'\Phi'-K\Phi^2+NK\Phi^2+rN\Phi^2K' =0,\nonumber\\
& &\nonumber\\
& &  Q^{(2)}{}_{\theta}= Q^{(3)}{}_{\phi}\equiv \Phi^2[\{2KN'+2NK'+r N'K'+2rNK''\}K-rNK'^2]+4KN\Phi\Phi'[2K+rK']\nonumber\\
&&\nonumber\\
& &+4rK^2[N'\Phi\Phi'-N\Phi'^2+2N\Phi\Phi'']=0, \nonumber\\
& & \nonumber\\
& &Q^{(4)}{}_t\equiv r\Phi^2N'+N\Phi^2+4rN\Phi\Phi'+r^2N'\Phi\Phi'-r^2N\Phi'^2+2r^2N\Phi\Phi''-\Phi^2=0, \nonumber\\
& &\nonumber\\
& & I\equiv 4rK^2N'\Phi-r^2NK'^2\Phi+2r^2NKK'' \Phi+r^2KN'K'\Phi+4NK^2\Phi-4K^2\Phi+4rNKK'\Phi+6r^2K^2N'\Phi'\nonumber\\
&&\nonumber\\
& &+24rNK^2\Phi'+6r^2NKK'\Phi'+12r^2NK^2\Phi''=0,
\end{eqnarray}
where $N'=\frac{dN(r)}{dr}$, $K'=\frac{dK(r)}{dr}$, $\Phi'=\frac{d\Phi(r)}{dr}$ and $\Phi''=\frac{d^2\Phi(r)}{dr^2}$.
The general solution of the above non-linear differential equations take the form:
\begin{eqnarray}
& &  \Phi(r)=\Phi(r),\qquad \qquad  K(r)=\frac{1}{\Phi^2(r)}+\frac{c_1}{r\Phi^3(r)}, \qquad \qquad N(r)=\frac{(r\Phi(r)+c_1)\Phi(r)}{r(\Phi(r)+r\Phi'(r))^2},
\end{eqnarray}
where $c_1$ is a constant of integration.

Now we are going to give a specific value of the scalar field $\Phi(r)$ and see what is the related values of the unknown $N(r)$ and $K(r)$.\vspace{0.2cm}\\ First of all, the value $\Phi(r)=\frac{1}{r}$, is not allowed since it makes the unknown function $N(r)$ has undefined  value. \vspace{0.2cm}\\ When $\Phi(r)=r$ we get
\begin{eqnarray} \label{sol}
K(r)=\frac{1}{r^2}+\frac{c_1}{r^4}, \qquad \qquad  N(r)=\frac{1}{4}+\frac{c_1}{4r^2}.
\end{eqnarray}
Using Eq. (\ref{met}) we get
\begin{eqnarray} \label{met1}
ds^2=\left(\frac{1}{r^2}+\frac{c_1}{r^4}\right)dt^2-\left(\frac{1}{4}+\frac{c_1}{4r^2}\right)^{-1}dr^2+r^2(d\theta^2+\sin^2\theta d\phi^2).\end{eqnarray}
Equation (\ref{met1}) shows that the above metric has no well-known asymptotic form, i.e., it has not a flat asymptote or de-Sitter/Anti-de-Sitter. Therefore we must exclude the choice $\Phi=r$.   \vspace{0.2cm}\\ If we continue in this manner we can show that  the only physical solution\footnote{We mean by physical solution is the one that has a well known asymptote behavior like the flat spacetime or AdS/dS spacetime.} of the above system can be read as:
\begin{eqnarray} \label{sol1}
& &  \Phi(r)=1,\qquad \qquad  N(r)=K(r)=1+\frac{c_1}{r}.
\end{eqnarray}
Calculating the metric of solution (\ref{sol1}) we get
\begin{eqnarray} \label{met2}
ds^2=\left(1+\frac{c_1}{r}\right)dt^2-\left(1+\frac{c_1}{r}\right)^{-1}dr^2+r^2(d\theta^2+\sin^2\theta d\phi^2).\end{eqnarray}
Equation (\ref{met2}) shows that the above metric has   a flat asymptote and is coincides with Schwarzschild spacetime provided that the constant  $c_1=-2m$ \cite{N06}. Therefore, we can conclude that for a diagonally static spherically symmetric spacetime the CTEGR theory in the vacuum case has no physical solution except the one known in TEGR which is the Schwarzschild solution. Is the structure of the vierbein given by Eq. (\ref{vierbein}) responsible for this result? In the next section we will use a non-diagonal form of a spherically symmetric spacetime and see if the scalar field $\Phi$ will have a non-constant value.
\section{Spherically symmetric black hole solution: Non-diagonal vierbein}\label{S3}

Now we apply the  field equations of CTEGR theory, Eqs. (\ref{q8}) and (\ref{q9}), to the non-diagonal static spherically symmetric $4$-dimensional  spacetime, written in the polar  coordinate ($r$, $\theta$, $\phi$, $t$) as follows \cite{Ngrg3,Nprd3}:
\begin{eqnarray}\label{vierbein1}
\nonumber \left({L^{i}}_{\mu}\right)=
\left( \begin{array}{cccc}
      \displaystyle \frac{\sin{\theta} \cos{\phi}}{\sqrt{N(r)}} & r\cos{\theta} \cos{\phi} & -r\sin\theta \sin{\phi}&0\\[7pt]
   \displaystyle\frac{\sin{\theta} \sin{\phi}}{\sqrt{N(r)}}& r\cos{\theta} \sin{\phi} & r\sin\theta \cos{\phi}&0\\[7pt]
   \displaystyle\frac{\cos{\theta}}{\sqrt{N(r)}}& -r\sin{\theta} & 0&0\\[7pt]
  0 & 0 & 0 &   \sqrt{K(r)} \\
  \end{array}
\right).\\
\end{eqnarray}
Using Eq. (\ref{vierbein1}) into Eq. (\ref{q33}) we get the non-vanishing components of the torsion $T^{a b c}$,  the vector of the torsion and contorsion $K^{a b c}$ in the form\footnote{The components of the torsion  $T^{a b c}$ is defined as \[T^{a b c}=L^a{}_\mu L^b{}_\nu L^c{}_\alpha T^{\mu \nu \alpha}.\] Same definition can be applied for the tensors $K^{\mu \nu \alpha}$ and $S^{\mu \nu \alpha}$.}:
\begin{eqnarray}\label{TC2}
 && T^{(1) (2) (1)}=T^{(3) (2) (3)}=\frac{\sin\theta \sin\phi(1-\sqrt{N})}{r},\qquad T^{(1) (3) (1)}=T^{(2) (3) (2)}=\frac{\cos\theta(1-\sqrt{N})}{r},\nonumber\\
&& T^{(2) (1) (2)}= T^{(3) (1) (3)}=\frac{\sin\theta \cos\phi(1-\sqrt{N})}{r}, \qquad  T^{(4) (1) (4)}=\frac{\sin\theta \cos\phi\sqrt{N}K'}{2K}, \nonumber\\
 &&T^{(4) (2) (4)}=\frac{\sin\theta \sin\phi\sqrt{N}K'}{2K},\qquad \qquad T^{(4) (3) (4)}=\frac{\cos\theta \sqrt{N}K'}{2K},  \nonumber\\
 && T_{(1)}=\frac{4K(1-\sqrt{N})-r\sqrt{N}K'}{2r\sqrt{N}K},\nonumber\\
& & K^{(2) (1) (1)}=K^{(2) (3) (3)}=\frac{\sin\theta \sin\phi(1-\sqrt{N})}{r},\qquad K^{ (3) (1) (1) }=K^{ (3) (2) (2) }=\frac{\cos\theta (1-\sqrt{N})}{r},\nonumber\\
& & K^{(1) (2)  (2)}=K^{(1) (3)  (3)}=\frac{\sin\theta \cos\phi(1-\sqrt{N})}{r},\qquad K^{(1) (4) (4)}=\frac{\sin\theta \cos\phi\sqrt{N}K'}{2K}, \qquad  K^{(2) (4) (4) }=\frac{\sin\theta \sin\phi\sqrt{N}K'}{2K},\nonumber\\
& & K^{(3) (4) (4) }=\frac{\cos\theta \sqrt{N}K'}{2K}.
\end{eqnarray}
Using Eq. (\ref{TC2}) we get the non-vanishing components of the superpotential in the form:
\begin{eqnarray}\label{sup1}
 && S^{ (1) (1)  (2)}=S^{ (3) (3) (2)}=\frac{\sin\theta\sin\phi(2K(1-\sqrt{N})+r\sqrt{N}K')}{4rK},\nonumber\\
& &  S^{ (2) (2) (1)}= S^{ (3) (3) (1)}=\frac{\sin\theta\cos\phi(2K(1-\sqrt{N})+r\sqrt{N}K')}{4rK},\nonumber\\
& & S^{ (2) (2)  (3)}=S^{ (1) (1)  (3)}=\frac{\cos\theta(2K(1-\sqrt{N})+r\sqrt{N}K')}{4rK}, \qquad S^{(4) (1) (4)  }=\frac{\sin\theta\cos\phi(K(1-\sqrt{N}))}{r}, \nonumber\\
& &   S^{(4) (2) (4)  }=\frac{\sin\theta\sin\phi(K(1-\sqrt{N}))}{r}, \qquad S^{(4) (3) (4)  }=\frac{\cos\theta(1-\sqrt{N})}{r}.
\end{eqnarray}
 Substituting from Eq. (\ref{vierbein}) into Eq. (\ref{Tor_sc}), we evaluate the torsion scalar as\footnote{The Ricci scalar of the spacetime (\ref{vierbein1}) is the same as given by Eq. (\ref{df1}).}.
\begin{equation}\label{df2}
T=\frac{2(1-\sqrt{N})(2K(1-\sqrt{N})+r\sqrt{N}K')}{r^2K}.
\end{equation}
Applying Eq. (\ref{vierbein1}) to the field equation (\ref{q8}) when ${\mathop{\mathfrak{T}}}^{\mu \nu}=0$ we get the same non-vanishing components as in the diagonal case.
Therefore, the same discussion applied  in the diagonal case can  be applied in this case.

As we see  form the above discussion that for the vacuum case and for a spherically symmetric spacetime either diagonal or  non-diagonal the scalar field $\Phi$ has no effect. Therefore, we will study the non-vacuum case in the following sections.

\section{A charged spherically symmetric black hole solution: diagonal vierbein}\label{S4}
Now we are going to derive the charged field equations of the CTEGR. For this aim,  we write the Lagrangian of the charged case in the form
\begin{equation}\label{q17}
\mathcal{L}({L^i}_\mu, \phi)_{g}=2\kappa|L|\left[-\phi^2T+6g^{\mu \nu}\partial_\mu \phi\partial_\nu \phi-4g^{\mu \nu}\phi(\partial_\nu \phi) T_\mu\right]+\mathcal{L}_{em},
\end{equation}
where $\mathcal{L}_{em}=-\frac{1}{2}{ F}\wedge ^{\star}{F}$ is the Maxwell Lagrangian,
with $F = dA$, and $A=A_{\mu}dx^\mu$, is the electromagnetic
potential 1-form.
Carrying out  the variational principle to the Lagrangian (\ref{q17}) with respect to the scalar field $\phi$ we get the same equation obtained in the non-charged case, i.e., Eq. (\ref{q8}) where   $\frac{\delta \mathcal{L}_m}{\delta \phi}=0$ since the trace of the energy-momentum tensor of the Maxwell field is vanishing
\begin{equation}\label{q88}
I\equiv \partial_\mu(Lg^{\mu \nu} \partial_\nu \phi)-\frac{L}{6}\phi R=0.\end{equation}   Now let us making  the variational principle of the Lagrangian density (\ref{q17}) with respect to the vierbein $L_{a \mu}$ we obtain \cite{MF}
\begin{eqnarray}\label{q19}
&&Q^{a \nu}\equiv \partial_\alpha(L\phi^2 S^{a \nu \alpha})-L\phi^2(S^{b  \alpha \nu} T_{b \alpha}{}^a-\frac{1}{4}L^{a \nu} T)
 -\frac{3}{2}LL^{a \nu}g^{\beta \mu}\partial_\beta \phi \partial_\mu \phi+3LL^{a \mu}g^{\beta \nu}\partial_\beta \phi \partial_\mu \phi
 +LL^{a \nu}g^{\beta \mu}  T_\mu \phi \partial_\beta \phi\nonumber\\
& & -L\phi L^{a \beta}g^{\nu \mu} ( T_\mu  \partial_\beta \phi+T_\beta  \partial_\mu \phi)
 -Lg^{\beta \mu}  \phi T^{\nu a}{}_\mu \partial_\beta \phi-\partial_\mu[L g^{\beta \nu} L^{a \mu} \phi \partial_\beta \phi]+\partial_\rho[L g^{\beta \rho} L^{a \nu} \phi \partial_\beta \phi]\nonumber\\
& &-\frac{\kappa \Phi^2}{2}(L^{a \rho}F_{\rho \alpha}F^{\nu \alpha}-\frac{1}{4}L^{a \nu}  F_{\alpha \beta}F^{\alpha \beta})=0.\end{eqnarray}
Finally, making the variational principle of the Lagrangian (\ref{q17}) with respect to the gauge potential $A$ we get
\begin{equation} \label{q29}
E^{\mu}\equiv {\rm \partial_\nu \left( \sqrt{-g} \Phi F^{\mu \nu} \right)}=0.
\end{equation}
It is important to note the fact that the field equations (\ref{q88}), (\ref{q19}) and (\ref{q29})  are reduced to the field equations of Maxwell-TEGR when the scalar field $\Phi=1$ \cite{N06,Nprd3}.

Applying the field equations (\ref{q88}), (\ref{q19}) and (\ref{q29})  to  the 4-dimensional spacetime of Eq. (\ref{vierbein}), after using Eqs. (\ref{TC1}),  (\ref{sup}), (\ref{df1}) and the following form of the gauge potential
\begin{equation}\label{df3} A = h(r) dt,\end{equation} we finally, get the following non-vanishing components:
\begin{eqnarray} \label{df4}
& & Q^{(1)}{}_r=3r^2NK\Phi'^2+4rNK\Phi \Phi'+r^2N\Phi K'\Phi'-K\Phi^2+NK\Phi^2+rN\Phi^2K'-4\pi r^2Nh'^2\Phi^2 =0,\nonumber\\
& &  \nonumber\\
& & Q^{(2)}{}_{\theta}= Q^{(3)}{}_{\phi}=\Phi^2[\{2KN'+2NK'+r N'K'+2rNK''\}K-rNK'^2]+4KN\Phi\Phi'[2K+rK']\nonumber\\
&&\nonumber\\
& & +4rK^2[N'\Phi\Phi'-N\Phi'^2+2N\Phi\Phi'']+16\pi rNK h'^2\Phi^2=0, \nonumber\\
& & \nonumber\\
& & Q^{(4)}{}_t=rK\Phi^2N'+NK\Phi^2+4rNK\Phi\Phi'+r^2KN'\Phi\Phi'-r^2KN\Phi'^2+2r^2KN\Phi\Phi''-K\Phi^2-4\pi r^2N h'^2\Phi^2=0,\nonumber\\
& & \nonumber\\
& & I= 4rK^2N'\Phi-r^2NK'^2\Phi+2r^2NKK'' \Phi+r^2KN'K'\Phi+4NK^2\Phi-4K^2\Phi+4rNKK'\Phi+6r^2K^2N'\Phi'\nonumber\\
&&\nonumber\\
& & +24rNK^2\Phi'+6r^2NKK'\Phi'+12r^2NK^2\Phi''=0,\nonumber\\
& & \nonumber\\
& & E^t\equiv 2rNKh' \Phi'-rNh'K'\Phi+rKN'h'\Phi+2rNKh'' \Phi+4NKh'\Phi=0,
\end{eqnarray}
where $h'=\frac{dh}{dr}$.

The above system of non-linear differential equation, Eq.  (\ref{df4}), reduces to the system of non-charged, Eq. (\ref{df11}), when the gauge potential $h(r)=1$. Also Eq. (\ref{df4}) reduces to the system of Maxewll-TEGR theory when $\Phi=1$ \cite{N06}. The  above differential equations, Eq. (\ref{df4}), have the following general solution
\newpage
\begin{eqnarray} \label{df6}
& &  \Phi(r)=\Phi(r), \qquad \qquad h(r)=c_2+c_3\int \frac{(r\Phi'+\Phi)}{r^2\Phi^3}dr,\nonumber\\
& & N(r)=\frac{c_4}{r^4\Phi'^2+2r^3\Phi\Phi'+r^2\Phi^2}\int\frac{r^4\Phi'^2+2r^3\Phi\Phi'+r^2\Phi^2}{r^2(r\Phi'+\Phi)}dr
+\frac{c_5}{r^2(r\Phi'+\Phi)^2}\nonumber\\
& &-\frac{2}{r^2(r\Phi'+\Phi)^2}\int \Bigg[(r\Phi'+\Phi)\Bigg(\int r\Phi'dr+\int \Phi dr\Bigg)\Bigg]dr-\Bigg(\int r\Phi'dr+\int \Phi dr\Bigg)\int (r\Phi'+\Phi)dr,\nonumber\\
& &
K(r)=-\frac{\int[2(r\Phi'+\Phi)(\int r\Phi'dr+\int \Phi dr)]dr-(2\int r\Phi'dr+2\int \Phi dr+c_3)\int (r\Phi'+\Phi)dr+c_5}{r^2\Phi^4[2\int( r\Phi'+\Phi)(\int r\Phi'dr+\int \Phi dr)dr+(2r\Phi -2\int r\Phi'dr-2 \int\Phi dr-c_4)\int(r\Phi'+\Phi)dr+\Phi^2r^2-r\Phi c_3+c_5]}.\nonumber\\
& &\end{eqnarray}
Equation  (\ref{df6}) shows that the solution of the non-linear differential equations given by Eq. (\ref{df4}) can not determine the unknown functions $\Phi$, $K(r)$, $N(r)$ and $h(r)$ in an explicate form. Therefore, we are going to see what is the best value of the arbitrary function $\Phi$ that makes the above solution is a physical one.

 Now let use assume $\Phi$ has the following value
\be \label{df7} \Phi(r)=\frac{1}{r},\ee
then using Eq. (\ref{df7}) in (\ref{df6}) we get
\be
 h(r)=c_2,\qquad \qquad  N(r)=undefined, \qquad \qquad K(r)=\frac{c_5r^2}{(1-c_3+c_4)}.\ee
 The above solution is not a physical one since the unknown function $N(r)$ is not defined therefore we must exclude the choice of $\Phi(r)$ given by Eq. (\ref{df7}).

 Now let use choice
\be \label{df8} \Phi(r)=r.\ee
Using Eq. (\ref{df8}) in (\ref{df6}) we get
\be \label{so5}
 h(r)=c_2-\frac{2c_2}{3r^3},\qquad \qquad  N(r)=\frac{1}{4}+\frac{c_3}{4r^2}+\frac{c_4}{4r^4}, \qquad \qquad K(r)=\frac{1}{c_3 r^2}+\frac{c_5}{c_3 r^4}+\frac{1}{r^6}.\ee
  Calculating the metric of solution (\ref{so5}) we get
\be \label{met3}
ds^2=\left(\frac{1}{c_3 r^2}+\frac{c_5}{c_3 r^4}+\frac{1}{r^6}\right)dt^2-\left(\frac{1}{4}+\frac{c_3}{4r^2}+\frac{c_4}{4r^4}\right)^{-1}dr^2+r^2(d\theta^2+\sin^2\theta d\phi^2).\ee
As we discussed before, equation (\ref{met3}) shows that the above metric has no well-known asymptotic form.  Therefore we must exclude also the choice $\Phi=r$.   \vspace{0.2cm}\\ Finally, if $\Phi=1$ then  the above system can be read as:
 \be \label{sol5} \Phi(r)=1, \qquad \qquad q(r)=c_2-\frac{c_3}{r}, \qquad \qquad N(r)=K(r)=1+\frac{c_4}{r}+\frac{c_3}{r^2}.\ee
Calculating the metric of solution (\ref{sol5}) we get
\be \label{met4}
ds^2=\left(1+\frac{c_4}{r}+\frac{c_3}{r^2}\right)dt^2-\left(1+\frac{c_4}{r}+\frac{c_3}{r^2}\right)^{-1}dr^2+r^2(d\theta^2+\sin^2\theta d\phi^2).\ee
As we discussed before, equation (\ref{met4}) shows that the above metric has   a flat asymptote and is coincides with Reissner-Nordstr\"{o}m spacetime provided that the constant  $c_4=-2m$ and $c_3=q$ \cite{N06}.

\section{A charged spherically symmetric black hole solution: Non-diagonal vierbein}\label{S5}

Applying the field equations (\ref{q88}), (\ref{q19}) and (\ref{q29})  to  the 4-dimensional spacetime of Eq. (\ref{vierbein1}), after using Eqs. (\ref{TC2}),  (\ref{sup1}), (\ref{df2}) and the gauge potential (\ref{df3}) we get the same non-linear differential equation (\ref{df4}). Therefore, the same discussion carried out in the charged diagonal vierbein can also be done here.

 \section{Conclusions}\label{S6}

In this study, we address the problem of finding spherically symmetric black hole solutions in the frame of CTEGR gravitational theory.  To aim this task, we apply the field equations of the CTEGR theory to a diagonally spherically symmetric spacetime.  We derive the full system of the non-linear differential equations. We derive an analytic solution of this system.  This analytic solution investigates in a clear way that the only physical solution that can be derived is achieved when the scalar field equal one, i.e., $\Phi(r)=1$.   This means that the conformal TEGR theory is identical with TEGR theory for the spherically symmetric spacetime.

To    give a concrete conclusion if these result  is always satisfied for the vacuum spherically symmetric spacetime we use a non-diagonal vierbein that has a spherically symmetric spacetime. We follow the same terminology used in the diagonal case and have derived the non-linear differential.  We have solved this system and have derived an analytic solution identical with the diagonal case. Therefore, we can conclude that for any spherically symmetric spacetime, the only vacuum black hole solution is identical with the TEGR solution.

To see if this result is always satisfied, we study non vacuum solution.  For this aim, we have derived the charged conformal TEGR field equations.  We  have applied these field equations to the diagonal spherically symmetric spacetime and have derived the system of differential equations.  We have solved this system and have derived an analytic solution. Like the vacuum case we have investigated in details that the only physical solution of this system is the Ressiner-Nordstr\"{o}m  spacetime with a trivial value of the scalar field $\Phi(r)=1$.   If we repeat the same procedure to the non-diagonal vierbein and apply the charged CTEGR to this vierbein we get the same results of the diagonal charged case. It is of interest to note that if we study the non-linear electrodynamics in the frame of CTEGR theory we will reach the same conclusion that the scalar field $\Phi$ will be trivial to get a physical solution.

Therefore, we can conclude that the only spherically symmetric  black hole solutions in the frame of CTEGR for non-charged and charged  are identical with the TEGR black hole solutions with trivial value of the scalar, i.e. $\Phi(r)=1$.  This indicates in a clear way that the scalar field $\Phi$ in the CTEGR theory has no effect for any spherically symmetric spacetime.

\subsection*{Acknowledgments}
 This work is partially supported by the Egyptian Ministry of Scientific Research under project No. 24-2-12. In addition, the work of KB is supported in part by the JSPS KAKENHI Grant Number JP 25800136 and Competitive Research Funds for Fukushima University Faculty (17RI017 and 18RI009).
\newpage


\begin{thebibliography}{99}

\bibitem{Ds} S. Deser,  {\it Ann. Phys.} {\bf 59} (1970), 248.

\bibitem{Dp} P. A. M. Dirac, {\it Proc. Roy. Soc. London} {\bf A 333} (1973), 403.

\bibitem{Ur} R. Utyiama, {\it Prog. Theor. Phys.} {\bf 53} (1975), 565.

 \bibitem{PA}  V. N. Pervushin, A. B. Arbuzov, B. M. Barbashov, R. G. Nazmitdinov, A. Borowiec,
K. N. Pichugin, and A. F. Zakharov, {\it Physics of Particles and Nuclei} {\bf 43} (2012), 682.

\bibitem{Ov}  V. I. Ogievetsky, {\it  Lett. Nuovo Cim.} {\bf 8} (1973), 988.

\bibitem{Hg}  G. \'{t} Hooft, {\it Found. Phys.} {\bf 41} (2011), 1829.
	
\bibitem{Hg10} G.  \'{t} Hooft, {\it arXiv:1009.0669}

\bibitem{Mp} P. D. Mannheim, {\it Prog. Part. Nucl. Phys.} {\bf 56} (2006), 340.

\bibitem{MS} T. Moon, P. Oh and J. Sohn, {\it JCAP} {\bf 1011} (2010), 005.

\bibitem{Ea28}    A. Einstein, {\it S.B. Preuss. Akad. Wiss.} (1928) 217- 221, see the translation in [arXiv:physics/0503046[physics.hist-ph]].

\bibitem{Nojiri:2010wj}
S.~Nojiri and S.~D.~Odintsov,
Phys.\ Rept.\  {\bf 505}, 59 (2011) 
[arXiv:1011.0544 [gr-qc]].

\bibitem{Capozziello:2011et}
S.~Capozziello and M.~De Laurentis,
Phys.\ Rept.\  {\bf 509}, 167 (2011) 
[arXiv:1108.6266 [gr-qc]].

\bibitem{Nojiri:2017ncd}
S.~Nojiri, S.~D.~Odintsov and V.~K.~Oikonomou,
Phys.\ Rept.\  {\bf 692}, 1 (2017) 
[arXiv:1705.11098 [gr-qc]].

\bibitem{Capozziello:2010zz}
V.~Faraoni and S.~Capozziello, 
Fundam.\ Theor.\ Phys.\  {\bf 170} (2010).

\bibitem{Bamba:2015uma}
  K.~Bamba and S.~D.~Odintsov,
  Symmetry {\bf 7}, 220 (2015)
  [arXiv:1503.00442 [hep-th]].

\bibitem{Bamba:2012cp}
K.~Bamba, S.~Capozziello, S.~Nojiri and S.D.~Odintsov, 
Astrophys.\ Space Sci.\  {\bf 342}, 155 (2012) 
[arXiv:1205.3421 [gr-qc]].

\bibitem{PP}  C. Pellegrini and J. Pleba\'{n}ski, {\it Mat.\ Fys.\
Scr.\ Dan.\ Vid.\ Selsk.\ }{\bf 2} (1963), no.4.

\bibitem{HS} K. Hayashi, {\it Phys. Lett.} {\bf 69B}, 441 (1977); K. Hayashi and T.
Shirafuji. {\it Phys. Rev.} {\bf D19}, 3524 (1979); {\it Phys.
Rev.} {\bf D24}, 3312 (1981); M. Blagojevi$\acute{c}$ and M.
Vasili$\acute{c}$  {\it Class. Quant. Grav.} {\bf 5} (1988), 1241;
T. Kawai, {\it Phys. Rev. } {\bf D62} (2000), 104014, T. Kawai, K.
Shibata and I. Tanaka, {\it Prog. Theor. Phys.} {\bf 104} (2000)
505.

\bibitem{Mj1} J. W. Maluf, Gen. Rel. Grav. 19, 57 (1987).

\bibitem{M1} J. W. Maluf, {\it J. Math. Phys.} {\bf 35 } (1994), 335.

\bibitem{MUFN} J. W. Maluf, S. C. Ulhoa, F. F. Faria, J.F. daRocha Neto, {\it Class.
Quant. Grav.} {\bf 23} (2006), 6245.

\bibitem{Mj} J. W. Maluf, {\it Ann.  Phys.} {\bf 14} (2005), 723.


\bibitem{MF} J.W. Maluf, F.F. Faria, {\it Ann. Phys.} {\bf 524} (2012), 366.

\bibitem{SSU} J. Silva, A. Santos and S. Ulhoa, {\it Eur. Phys. J.} {\bf C 76} (2016), 1.

\bibitem{Hawking:1972qk} 
  S.~W.~Hawking,
  Commun.\ Math.\ Phys.\  {\bf 25}, 167 (1972).
  doi:10.1007/BF01877518

\bibitem{Mayo:1996mv} 
  A.~E.~Mayo and J.~D.~Bekenstein,
  Phys.\ Rev.\ D {\bf 54}, 5059 (1996)
  [gr-qc/9602057].

\bibitem{Bekenstein:1996pn} 
  J.~D.~Bekenstein,
  In *Moscow 1996, 2nd International A.D. Sakharov Conference on physics* 216-219
  [gr-qc/9605059].

\bibitem{Sotiriou:2011dz} 
  T.~P.~Sotiriou and V.~Faraoni,
  Phys.\ Rev.\ Lett.\  {\bf 108}, 081103 (2012)
  [arXiv:1109.6324 [gr-qc]].

\bibitem{Bhattacharya:2015iha} 
  S.~Bhattacharya, K.~F.~Dialektopoulos, A.~E.~Romano and T.~N.~Tomaras,
  Phys.\ Rev.\ Lett.\  {\bf 115}, 181104 (2015)
  [arXiv:1505.02375 [gr-qc]].

\bibitem{Wr} R. Weitzenb\"ock, {\it Invariance Theorie}, Nordhoff, Groningen, 1923.


\bibitem{AP} R. Aldrovandi and J. G. Pereira, \textit{Teleparallel Gravity: An Introduction}, Springer, Dordrecth,
(2012), \textit{Teleparallel Gravity} at http://www.ift.unesp.br/users/jpereira/tele.pdf.

\bibitem{MDTC} J. W. Maluf, J. F. da Rocha-neto, T. M. L. Toribio and K. H.
Castello-Branco,  {\it Phys.\ Rev.\ }{\bf D65} (2002), 124001.

\bibitem{US3} S. C. Ulhoa and E. P. Spaniol,  {\it Int J. Mod. Phys. } {\bf D22} (2013), 1350069.

\bibitem{N008}  G. G. L. Nashed, {\it Euro. Phys. J.} {\bf C 54} (2008), 2,  291 [ arXiv:0804.3285 [gr-qc]].

\bibitem{N010}  G. G. L. Nashed, {\it Astrophys. Space Sci.} {\bf 330} (2010), 173 [arXiv:1503.01379 [gr-qc]].

\bibitem{Gw10}  G. G. L. Nashed, {\it Chin. Phys.} {\bf B 19} (2010), 2,  020401 [arXiv:0910.5124  [gr-qc]].

\bibitem{N06}  G. G. L. Nashed, {\it Mod. Phys. Lett.} {\bf A 21} (2006), 29,  2241 [  arXiv:gr-qc/0401041 [gr-qc]].

\bibitem{Ngrg3}  G. G. L. Nashed, {\it Gen. Rel. Grav.}  {\bf 45} (2013), 1887.

\bibitem{Nprd3} G. G. L. Nashed {\it  Phys. Rev.} {\bf D 88} (2013),  104034.



\end{thebibliography}
\end{document}